\documentclass{iucrjournals}

\nolinenumbers

\usepackage{amsmath}
\usepackage{amssymb}
\usepackage{graphicx}   
\usepackage{makecell}
\usepackage{array}
\usepackage{makecell}
\usepackage{endnotes}
\usepackage{caption}
\usepackage{tabularx}

\title{Modified Mosseri-Sadoc tiles from $D_6$}
 
\author[a]{Rehab Al Raisi}%
\author[a,*]{Nazife Ozdes Koca}%
\author[a,**]{Mehmet Koca}%
\author[b]{Ramazan Koc}

\affil[a]{Department of Physics, College of Science, Sultan Qaboos University,  P.O. Box 36, Al-Khoud 123, Muscat, Sultanate of Oman,}
\affil[b]{Department of Physics, Gaziantep University, Gaziantep, Turkey}
\affil[**]{Professor Emeritus}
\affil[*]{Correspondence e-mail: nazife@squ.edu.om}

\begin{document}
\maketitle
\begin{center}
{\Large \textbf{Abstract}}  
\end{center}
A modified set of Mosseri-Sadoc (MS) tiles tessellating 3D Euclidean space with icosahedral  symmetry is introduced. The new set of tiles are embedded in dodecahedron with a threefold symmetric order. The modified Mosseri-Sadoc (MMS) tiles can be inflated by a new inflation matrix with positive eigenvalues $\tau^3$ and $\tau$ with the corresponding eigenvectors representing the volumes and the Dehn invariants of the tiles, respectively, where $\tau=\frac{1+\sqrt5}{2}$ is the golden ratio. The MMS tiles are obtained by projection of the 4D and 5D facets of the Delone cells tiling the $D_6$ root lattice in an alternating order. It is also proved that a subset of the lattice $D_6$ projects into the dodecahedron inflated by $\tau^n$ with an arbitrary integer $n$ and tiled by the MMS tiles.
\newline\newline
\noindent
\textbf{Keywords:} Icosahedral quasicrystals, aperiodic tiling, lattices, projections of polytopes, polyhedra.
\newpage


\section{Introduction}

The discovery of icosahedral symmetric quasicrystals \cite{Shechtman1984}, along with many planar quasicrystalline structures exhibiting fivefold, eightfold, tenfold, twelvefold, and eighteenfold symmetries, has led to numerous theoretical developments, including lattice projection models. For a general exposition, the reader is referred to the references on quasicrystallography \cite{DiVincenzo1991,Janot1993,Senechal1995,Steurer2004,Tsai1994,Tsai2008}. Among the lattice projection models the checkerboard root lattice $D_6$\ \cite{Conway1991,Conway1999} has been the center of focus in the description of the icosahedral symmetric quasicrystallography for the icosahedral Coxeter group $H_3$ is a maximal subgroup of $D_6$. Several attempts have proved that the $D_6$ lattice provides a useful framework for the icosahedral symmetric tiling scheme. As early as 1982, Kramer \cite{Kramer1982} has introduced a seven-tile system which was first reduced to a six-tetrahedral tile model leading to the MS four-tile system \cite{Mosseri1982}. Importance of the $D_6$ lattice is verified by Kramer and Papadopolos \cite{Kramer1994} by deriving the six-tetrahedral tile system from the projection of the root lattice $D_6$\ employing the cut-and-project technique. Later it was demonstrated \cite{Papadopolos2000} that the six-tetrahedral tiling system does not admit stone inflation due to Dehn invariance. There it was also shown that the Mosseri-Sadoc $4 \times 4$ inflation matrix can be obtained under the assumption that the eigenvectors of the inflation matrix correspond to the volumes and Dehn invariants of the tiles, with positive eigenvalues $\tau^3$ and $\tau$, respectively.\newline\noindent\hspace*{0.6cm}
Recently, it was suggested that \cite{Koca2020,Koca2021} the six-tetrahedral tile system can be obtained by projecting the 3D facets of the Delone cells of the root lattice $D_6$  into what is usually called the $E_\parallel$ space. The tetrahedral tiles are classified as the fundamental tiles and denoted by $\tau_i$ $(i=1,2,…,6)$. The fundamental tiles represent the tetrahedra consisting of 1, 2, 3, 3, 4, 5 edges of length $\tau$ for the tiles $\tau_1$, $\tau_2$, $\tau_3$, $\tau_4$, $\tau_5$,$\tau_6$, respectively, and the remaining edges are of  unit length. The MS four-tile system is obtained by gluing the equilateral faces of the tetrahedral tiles, as these faces cannot be dissected into similar ones under golden ratio inflation. The MS tiling system is not unique in tiling of the 3D Euclidean space with icosahedral symmetry; there exist a set of four prototiles \cite{Socolar1986} consisting of acute rhombohedron, Bilinski dodecahedron, rhombic icosahedron and rhombic triacontahedron. A decorated version of the Ammann tiles were proposed by Katz \cite{Katz1989} and it has been revived by Hann-Socolar-Steinhardt \cite{Hann2016}. Danzer \cite{Danzer1989} proposed a more fundamental tiling scheme known as the $ABCK$ tetrahedral prototiles with its octahedral tiling is denoted by $\langle ABCK \rangle$. It has been shown later that the Ammann tiles and the Danzer tiles follow from each other \cite{Danzer1993,Roth1993}.\newline\noindent\hspace*{0.6cm}
It was further shown that the Socolar-Steinhardt tiles arise from projections of the lattice sequence $B_3\subset B_4\subset B_5\subset B_6$, with each lattice producing, under projection, an acute rhombohedron, a Bilinski dodecahedron, a rhombic icosahedron, and a rhombic triacontahedron, respectively \cite{Koca2015}. These facts indicate that the icosahedral symmetric tiling systems may follow from the lattice projections of $D_6\subset B_6$ system where $AutD_6\approx B_6$. Not only icosahedral symmetric quasicrystallography but also the $h$-fold symmetric planar quasicrystallography may also follow from the $A_n$ lattice projections \cite{Koca2019} where $h=n+1$ is the Coxeter number.\newline\noindent\hspace*{0.6cm}
In this paper we demonstrate that the MS four-tile system can be modified possessing a deeper group theoretical meaning and allowing a threefold symmetric embedding of the tiles in dodecahedron, the unique icosahedral symmetric polyhedron tiled by the MMS tiles. We show that the six-tetrahedral tiles follow from the projections of 3D facets of the Delone cells of the root lattice $D_6$ without invoking the cut-and-project technique. Furthermore, the MS tiles and MMS tiles can be obtained from projections of the 4D and 5D facets of the Delone cells of $D_6$. Since the Delone cells tile the root lattice in an alternating order \cite{Conway1999} it is expected that the tiles projected from the Delone cells may tile the 3D Euclidean space in a similar manner with an icosahedral symmetry. The weight vectors $\omega_i\ (i=1,...,6)$ of the root lattice $D_6$ determine certain polytopes in 6D Euclidean space among which the union of  orbits of the weights $\omega_1,\ \omega_5$ and $\omega_6$ represent the Voronoi cell $V\left(0\right)$ centered around the origin and the orbits of the same weights constitute the Delone cells according to the rule:${W(d}_{6})\omega_1+{W(d}_{6})\omega_1, {{W(d}_{6})\omega}_5+{{W(d}_{6})\omega}_5$ and${{ W(d}_{6})\omega}_6+{{W(d}_{6})\omega}_6$ where ${ W(d}_{6})$ is the Coxeter-Weyl group of order $2^56!$ generated by reflections. Here, with ${{W(d}_{6 })\omega}_6+{{W(d}_{6})\omega}_6$, for example, it is meant that the vector sum must be taken to construct the Delone cells whose centroids constitute the part of the vertices of the Voronoi cell $V\left(0\right)$. \newline\noindent\hspace*{0.6cm}
The paper is organized as follows. In Sec. 2, we introduce a short review on the structure of the point group $W\left(d_6\right)$ in the framework of the Coxeter-Dynkin diagram and its mapping into the 3D Euclidean space and revise the six-tetrahedral tile system (fundamental tiles ${t}_i)$ projected from 3D facets of the Delone cells of the root lattice $D_6$. We also construct the subset of the lattice $D_6$ which projects into dodecahedron inflated, in an arbitrary integer order, by $\tau$. In Sec. 3 four MMS tiles are assembled using the MS tiles obtained from the fundamental tiles $t_i$ so that their faces consist of the Robinson triangles, trapezoids and pentagons only normal to the fivefold axes. Volume vector and the Dehn vector of the modified tiles are constructed to determine the new $4 \times 4$ inflation matrix $N$ whose eigenvalues are\ $\tau^3, \tau,\sigma$ and $\sigma^3$ where $\sigma=-\ \tau^{-1}=\frac{1-\sqrt5}{2}$ is the algebraic conjugate of the golden ratio $\tau$. The right and left eigenvectors of the inflation matrix $N$ corresponding to the Perron-Frobenius (PF) eigenvalue $\tau^3$ are calculated and the projection matrix is formed as the tensor product of the right and left eigenvectors. In Sec. 4 we study the structures of the 4D and 5D facets of the Delone cells and show how the MMS tiles follow from projections of the 4D and 5D facets of the Delone cells. In Sec. 5 we discuss the group theoretical structure leading to the threefold symmetric tiling of dodecahedron in terms of the MMS tiles. Concluding remarks include a general symmetry framework of the possible tiles arising from the sequence of quasicrystallographic groups $H_2\subset H_3\subset H_4$ embedded in the respective point groups of the lattices $A_4 \subset D_6\subset E_8$. Appendix A (Table 3) includes the list of facets of the orbits of the fundamental weights $\omega_i\,  \left(i=1,\ldots,6\right)$.

\section{Diagrammatic representation of \texorpdfstring{$D_6$ and $H_3$}{D6 and H3}}
In this section, we determine a subset of the root lattice $D_6$ that projects onto a dodecahedron inflated by an arbitrary integer power of the golden ratio $\tau$. Using the Coxeter-Dynkin diagrams of $d_6$ and $h_3$ (Fig. 1), we introduce the roots, weights, and the projection technique. An arbitrary vector of the root lattice $D_{6\ }$ can be written as a linear combination of the simple roots with integer coefficients,
\begin{equation}
\lambda=\sum_{i=1}^{6} n_i \alpha_i = \sum_{i=1}^{6} m_i l_i,\quad  \sum_{i=1}^{6} m_i \text{ is even}, \quad
n_i, m_i \in \mathbb{Z}
\end{equation}
where $\alpha_i$ are the simple roots of $D_6$ defined in terms of the orthonormal set of vectors $l_i\left(i=1,2,\ldots,6\right)$ as $\alpha_i=l_i-l_{i+1},i=1,…,5$ and $\alpha_6=l_5+l_6$. The reflection generators of ${W(d}_{6\ })$ act on the orthonormal set of vectors as $r_i:l_i\longleftrightarrow l_{i+1}\ and\ r_6:l_5\longleftrightarrow-l_6$. Then the weight vectors of ${W(d}_{6})$ defined by the relation $\left(\alpha_i,\omega_j\right)=\delta_{ij}$ are given by,
\begin{equation}
\begin{aligned}
\omega_1 &= l_1, \\
\omega_2 &= l_1 + l_2, \\
\omega_3 &= l_1 + l_2 + l_3, \\
\omega_4 &= l_1 + l_2 + l_3 + l_4, \\
\omega_5 &= \tfrac{1}{2}\left(l_1 + l_2 + l_3 + l_4 + l_5 - l_6\right), \\
\omega_6 &= \tfrac{1}{2}\left(l_1 + l_2 + l_3 + l_4 + l_5 + l_6\right).
\end{aligned}
\end{equation}\noindent\hspace*{0.6cm}
The orbits ${W(d}_{6})\omega_1, {W(d}_{6})\omega_5$ and ${W(d}_{6})\omega_6$ are the Delone cells tiling the root lattice in an alternating order as expressed in the introductory section. For example, the orbit ${W(d}_{6 })\omega_1=\pm l_i,\left(i=1,\ 2,\ldots,6\right),$ represents an orthoplex, the octahedron in 6D. Similarly, the orbits ${W(d}_{6})\omega_5$ and ${W(d}_{6})\omega_6$ represent two 6D hemicubes whose union is a 6D cube. The Delone cells whose centroids constitute the vertices of the Voronoi cell $V\left(0\right)$ of the root lattice $D_6$ can be obtained as the vector sum of the orbits ${\ W(d}_{6})\omega_i+{W(d}_{6})\omega_i,\ (i=1,\ 5,\ 6)$ which constitute the vectors of norm 0, 2, 4, 6.\newline\noindent\hspace*{0.6cm}
The generators of the icosahedral subgroup ${W(h}_{3})=<R_1,R_2,R_3>$ of ${W(d}_{6})$ can be defined as \cite{Koca2015}, $R_1=r_1r_5,R_2=r_2r_4$ and $R_3=r_3r_6$ where the Coxeter element \cite{Coxeter1973} for example, can be taken as $R=R_1R_2R_3$ with $R^{10}=1$. The simple roots and weights of the group $W(h_{3})$ are defined in two complementary 3D Euclidean spaces $E_\parallel$ and $E_\perp{\ \beta}_i({\acute{\beta}}_i)$ and $v_i\left({\acute{v}}_i\right),\left(i=1,2,3\right)$ respectively, then they can be expressed in terms of the roots and weights of $W(d_{6})$ as
\begin{equation}
\begin{aligned}
{\ \beta}_1=\frac{1}{\sqrt{2+\tau}}\left(\alpha_1+\tau\alpha_5\right),{\ v}_1=\frac{1}{\sqrt{2+\tau}}\left(\omega_1+\tau\omega_5\right),\\
{\ \beta}_2=\frac{1}{\sqrt{2+\tau}}\left(\alpha_2+\tau\alpha_4\right), {\ v}_2=\frac{1}{\sqrt{2+\tau}}\left(\omega_2+\tau\omega_4\right),  \\    
{\ \beta}_3=\frac{1}{\sqrt{2+\tau}}\left(\alpha_6+\tau\alpha_3\right),{\ v}_3=\frac{1}{\sqrt{2+\tau}}\left(\omega_6+\tau\omega_3\right).
\end{aligned}
\end{equation}
For the complementary 3D Euclidean space replace $\beta_i$ by ${\acute{\beta}}_i$, $v_{i\ }$ by ${\acute{v}}_i$ and $\tau$ by $\sigma$  in (3). \newline
Fig.1 demonstrates the decomposition of the 6D space into a direct sum of two complementary 3D Euclidean spaces, in which the nodes $\alpha_i,\left(i=1,2,\ldots,6\right), \beta_i,\left(i=1,2,3\right)$ represent the associated simple roots, and ${\acute{\beta}}_i$ corresponds to $\beta_i$ in the complementary space.
\begin{figure}[ht] %
\label{fig:figure1}
\begin{center}
\includegraphics[width=0.8\textwidth]{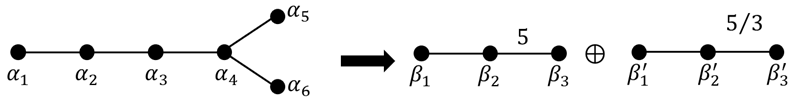} 
\end{center}
\caption{The Coxeter-Dynkin diagrams of $d_6$ and $h_3$ illustrating the symbolic projection.}
\end{figure}
For a choice of the simple roots of $h_3$ as $\beta_1=\left(\sqrt2,0,\ 0\right),\beta_2=-\frac{1}{\sqrt2}\left(1,\sigma,\tau\right),\beta_3=\left(0,\ 0,\sqrt2\right)$ and the similar expressions for the roots in the complementary space we can express the components of the set of vectors $l_i \left(i=1,2,\ldots,6\right)$ as
\begin{equation}
\begin{aligned}
\left[\begin{matrix}l_1\\l_2\\\begin{matrix}l_3\\\begin{matrix}l_4\\\begin{matrix}l_5\\l_6\end{matrix}\end{matrix}\end{matrix}\end{matrix}\right]=\frac{1}{\sqrt{2\left(2+\tau\right)}}\left[\begin{matrix}\ \ \ 1&\tau&\ \ \ 0&\ \ \ \tau&-1&\ \ \ 0\\-1&\tau&\ \ \ 0&-\tau&-1&\ \ \ 0\\\ \ \ 0&1&\ \ \ \tau&\ \ \ 0&\ \ \ \tau&-1\\\ \ \ 0&1&-\tau&\ \ \ 0&\ \ \ \tau&\ \ \ 1\\\ \ \ \tau&0&\ \ \ 1&-1&\ \ \ 0&\ \ \ \tau\\-\tau&0&\ \ \ 1&\ \ \ 1&\ \ \ 0&\ \ \ \tau\end{matrix}\right].
\end{aligned}
\end{equation}
First three and last three components project the vectors $l_i$ into $E_\parallel$ and $E_\bot$ spaces, respectively. On the other hand, the weights of $h_3$ are represented by the vectors $v_1=\frac{1}{\sqrt2}\left(1,\tau,0\right),\ v_2=\sqrt2\left(0,\tau,0\right),v_3=\frac{1}{\sqrt2}\left(0,\tau^2,1\right)$. Note that the vector $v_3$ is invariant under the dihedral group $W(a_{2})$, of order 6, generated by the two generators $R_1=r_1r_5$ and $R_2=r_2r_4$. This feature of dodecahedron allows the modification of the MS tiles and impose the threefold symmetric embedding of the tiles into  dodecahedron. The orbits ${W(h}_{3\ })v_i,\left(i=1,2,3\right)$ represent the sets of vertices of an icosahedron, an icosidodecahedron and a dodecahedron respectively where the vertices of the dodecahedron is shown explicitly as
\begin{equation}
\begin{aligned}
{W(h}_{3})v_3=\frac{\tau}{\sqrt2}\left\{\left(\pm1,\pm1,\pm1\right),\left(0,\pm\tau,\pm\sigma\right),\left(\pm\tau,\pm\sigma,0\right),\left(\pm\sigma,0,\pm\tau\right)\right\}.
\end{aligned}
\end{equation}
A general lattice vector can be written as a linear combination of the weights ${ v}_i\left({\acute{v}}_i\right)$ as 
\begin{equation}
\begin{aligned}
m_1 l_1 + m_2 l_2 + m_3 l_3 + m_4 l_4 + m_5 l_5 + m_6 l_6
&= \frac{1}{\sqrt{2+\tau}} \Bigl[
    \bigl(m_1 - m_2 + \tau m_5 - \tau m_6\bigr) v_1 \\
&\quad + \bigl(m_2 - m_3 + \tau m_4 - \tau m_5\bigr) v_2 \\
&\quad + \bigl(m_5 + m_6 + \tau m_3 - \tau m_4\bigr) v_3
\Bigr] \\
&\quad + \text{($v_i \longrightarrow \acute{v}_i$, $\tau \longrightarrow \sigma$)}.
\end{aligned}
\end{equation}
The lattice vector projecting on the vector representing the dodecahedron can be expressed as
follows \cite{AlSiyabi2020}:
\begin{equation}
\begin{aligned}
\left[m_1\left(l_1+l_2+l_3\right)+m_4\left(l_4+l_5+l_6\right)\right]_\parallel=\frac{1}{\sqrt{2+\tau}}\left(2m_4+\tau m_1-\tau m_4\right){v}_3.
\end{aligned}
\end{equation}
Applying the generators of the $W(h_{3})$ one can prove that the vertices of the dodecahedron can be obtained in terms of the pair of integers $\left(m_1,m_4\right)$ being either both odd or both even integers. This subset of lattice vectors defines the concentric dodecahedra with varying radii of circumference. A general lattice vector $\sum_{i=1}^{6}{n_il_i}$ will shift the center of dodecahedron to an arbitrary point of the lattice space. An inflation of dodecahedron $W(h_{3})v_3$ by $\tau^n$ replaces the integers in (7) by the pair of integers $\left(\acute{m_1,}\acute{m_4}\right)$ which can be proven that they are either both even or both odd integers indicating that the inflated dodecahedra remain in the same subset of the lattice. This is also true for the icosahedral symmetric Platonic and Archimedean polyhedra; but our main concern here is that the MMS tiles, that we are going to discuss in this paper, tile only inflated dodecahedra in threefold symmetric manner. It is because dodecahedron is the unique polyhedron consisting of pentagonal faces only. \newline\noindent\hspace*{0.6cm}
Other icosahedral symmetric polyhedra also possess, in addition to pentagonal or decagonal faces, triangular or hexagonal faces that are incompatible with MMS tiles.


\section{Modified Mosseri-Sadoc (MMS) tiles}
Earlier \cite{Koca2020,Koca2021} we have obtained the 6 fundamental tetrahedral tiles $t_i\left(i=1,2,\ldots,6\right)$ as the projections of the 1520 regular 3D tetrahedral facets of the Delone cells of the lattice $D_6$. The projected fundamental tiles are the tetrahedra with faces of equilateral and Robinson triangles. The fundamental tiles $t_i$ have 1, 2, 3, 3, 4, 5 edges of length $\tau$, respectively. Similarly, their faces consist of equilateral triangles arranged as follows: 2 triangles of edge length 1, 1 triangle of edge length 1, 1 triangle of edge length $\tau$, 1 triangle of edge length 1, 1 triangle of edge length $\tau$, and 2 triangles of edge length $\tau$, respectively, as shown in Table 1.
\newpage
\begin{table}[ht]
\caption{Fundamental tiles projected from 3D facets of the Delone cells $\omega_1$, $\omega_5$ and $\omega_6$.}
\begin{center}
\begin{tabular}{c m{2.5cm} c c c c c}
\midrule
\makecell{Fundamental\\tiles} & Figure & \makecell{Number of\\faces} & Volume & Count in $\omega_1$ & Count in $\omega_5$ & Count in $\omega_6$ \\
\midrule
$t_1$ & \includegraphics[width=2cm]{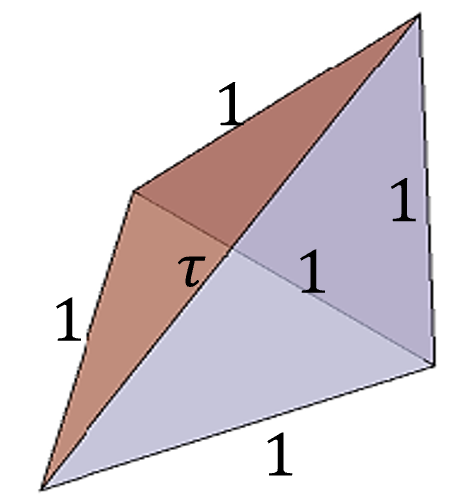} & \makecell{$2 \times \{1,1,1\}$\\$2 \times \{1,1,\tau\}$}
& $\frac{1}{12}$ & 30 & 60 & 60 \\
$t_2$ & \includegraphics[width=2cm]{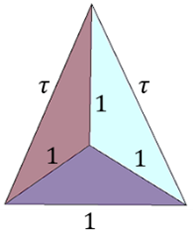} & \makecell{$1 \times \{1,1,1\}$\\$2 \times \{1,1,\tau\}$\\$2 \times \{1,\tau\,\tau\}$}
& $\frac{\tau}{12}$ & 60 & 120 & 120 \\
$t_3$ & \includegraphics[width=2cm]{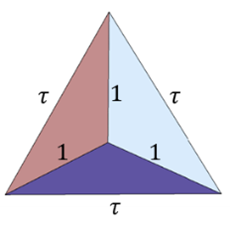} & \makecell{$1 \times \{\tau,\tau\,\tau\}$\\$3 \times \{1,1,\tau\}$}
& $\frac{\tau}{12}$ & - & 80 & 80  \\
$t_4$ & \includegraphics[width=2cm]{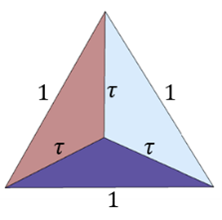} & \makecell{$1 \times \{1,1,1\}$\\$3 \times \{1,\tau,\tau\}$}
& $\frac{\tau^2}{12}$ & - & 80 & 80 \\
$t_5$ & \includegraphics[width=2cm]{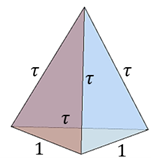} & \makecell{$1 \times \{\tau\,\tau\,\tau\}$\\$1 \times \{1,1,\tau\}$\\$2 \times \{1,\tau\,\tau\}$} & $\frac{\tau^2}{12}$ & 60 & 120 & 120 \\
$t_6$ & \includegraphics[width=2cm]{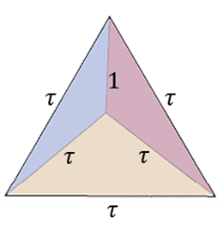} & \makecell{$2 \times \{\tau\,\tau\,\tau\}$\\$2 \times \{1,\tau,\tau\}$} & $\frac{\tau^3}{12}$ & 30 & 60 & 60 \\ \hline
\end{tabular}
\end{center}
\end{table}
\noindent\hspace*{0.6cm}
The other faces are the Robinson triangles of edge lengths $\left(1,\tau,1\right)$ and $\left(\tau,1,\tau\right)$. Since the inflated equilateral triangular faces by $\tau$ cannot be dissected into smaller triangles of the same kind, the composite tiles (MS tiles) are constructed as the tiles by gluing the corresponding equilateral triangles of the fundamental tiles. The original notation of the MS tiles $H, A, S, Z$ \cite{Mosseri1982} are then renamed as  ${\ T}_1,{\ T}_2,{\ T}_3,{\ T}_4$, respectively, which were composed of the fundamental tiles as follows:     
\begin{equation}
\begin{alignedat}{1}
T_1 &= (t_4 + t_1 + t_4) + (t_3 + t_6 + t_3),\\
T_2 &= t_2 + t_4,\\
T_3 &= t_5 + t_6 + t_5,\\
T_4 &= t_5 + t_6 + t_3.
\end{alignedat}
\end{equation}
\noindent\hspace*{0.6cm}
Details of this construction can be found in reference \cite{Koca2021}. See also \cite{Papadopolos2000}. A natural question arises as to whether a new set of composite tiles can be obtained by reshuffling the MS tiles. This question is motivated by the tiling of the dodecahedron with MS tiles, particularly due to the role of the vector ${\pm\ v}_3$ is invariant under the dihedral group ${W(a}_{2\ })= <R_1,R_2>$, of order 6, with ${R_1}^2={R_2}^2=\left(R_1R_2\right)^3=1$. This implies that the dodecahedron can be tiled in a threefold symmetric manner. Indeed, this is possible by redefining the MMS tiles as follows:
\begin{equation}
\begin{alignedat}{1}
\bar{T_1}&=: T_1+ T_2+T_4 ,\\
\bar{T_2}&=: T_2\\
\bar{T_3}&=: T_3\\
\bar{T_4}&=: T_2+T_4.                              
\end{alignedat}
\end{equation}
\noindent\hspace*{0.6cm}
Faces of the MMS tiles are all either Robinson triangles, trapezoids or pentagons which can be further dissected into Robinson triangles. The MS tiles, contrary to the MMS tiles, do not have pentagonal face. It is worth noting that tilings obtained by projection of the Delone cells of the  root lattice $A_4$ onto the Coxeter plane contains only the Robinson triangles of two types with edges $\left(1,\tau,1\right)$ and $\left(\tau,1,\tau\right)$ \cite{Koca2019}. The Robinson triangles admit an inflation rule and the Penrose dart and kite tiles can be obtained as composite tiles by gluing together the corresponding edges of Robinson triangles. Since $A_4$ is the sublattice of the lattice $D_6$ it is quite natural to expect that the MS or MMS tiles should have faces containing Robinson triangles. While constructing the MMS tiles, one must ensure that like faces are matched: Robinson triangles with Robinson triangles, trapezoids with trapezoids, and pentagons with pentagons. Number of vertices, $N_0$, edges, $N_1$, and faces, $N_2$, including the volumes and Dehn invariants of the MMS tiles are displayed in Table 2. The Dehn invariant of a polyhedron P is defined by the equation
\begin{equation}
\begin{alignedat}{1}
\mathcal{D}\left(P\right)=\sum{l_i}\otimes{\bar{\alpha}_i},
\end{alignedat}
\end{equation}
where $l_i$ denotes the length of edges of P and  ${\bar{\alpha}}_i$ is the dihedral angle modulo $\pi$. In the composite tiles the dihedral angles are either $\tan^{-1}{(2})$ or $\pi-\tan^{-1}{\left(2\right)}$. \newline
The Mosseri-Sadoc inflation matrix 
\begin{equation}
\begin{alignedat}{1}
M=\left(\begin{matrix}\begin{matrix}1\\\begin{matrix}0\\\begin{matrix}1\\1\end{matrix}\end{matrix}\end{matrix}&\begin{matrix}\begin{matrix}2\\\begin{matrix}2\\\begin{matrix}2\\1\end{matrix}\end{matrix}\end{matrix}&\begin{matrix}\begin{matrix}2\\\begin{matrix}1\\\begin{matrix}1\\1\end{matrix}\end{matrix}\end{matrix}&\begin{matrix}2\\\begin{matrix}0\\\begin{matrix}1\\1\end{matrix}\end{matrix}\end{matrix}\end{matrix}\end{matrix}\end{matrix}\right),
\end{alignedat}
\end{equation}
can be modified for MMS tiles as 
\begin{equation}
\begin{alignedat}{1}
N=\left(\begin{matrix}\begin{matrix}2\\\begin{matrix}0\\\begin{matrix}1\\1\end{matrix}\end{matrix}\end{matrix}&\begin{matrix}\begin{matrix}2\\\begin{matrix}2\\\begin{matrix}1\\2\end{matrix}\end{matrix}\end{matrix}&\begin{matrix}\begin{matrix}4\\\begin{matrix}1\\\begin{matrix}1\\2\end{matrix}\end{matrix}\end{matrix}&\begin{matrix}1\\\begin{matrix}0\\\begin{matrix}0\\0\end{matrix}\end{matrix}\end{matrix}\end{matrix}\end{matrix}\end{matrix}\right),
\end{alignedat}
\end{equation}
which can be obtained from the matrix $M$ by a similarity transformation based on the relation (9).
Therefore, the eigenvalues of the matrix $N$ are $\tau^3$, $\tau$, $\sigma$ and $\sigma^3$ as those of the matrix $M$ and the eigenvectors corresponding to $\tau^3$ and $\tau$ are the volume vector and the Dehn invariant vector listed in Table 2.
\begin{table}[ht]
\caption{Some properties of the tiles $\bar{T_1}, \bar{T_2} ,\bar{T_3}, \bar{T_4}$ where $N_0$ is the number of vertices, $N_1$ is the number of edges, and $N_2$ is the number of faces.}
\smallskip
\begin{center}
\begin{tabular}{c m{2.5cm} c c c c c c}
\midrule
\makecell{Composite\\tiles} & Figure & $N_0$ & $N_1$ & $N_2$ & Types of faces & Volume & \makecell{Dehn\\Invariant}  \\
\midrule
$\bar{T_1}$ & \includegraphics[width=2cm]{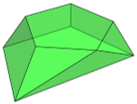} & 9 & 15 & 8 & 
\makecell{
$1 \times \{\tau,\tau,\tau,\tau^2\}$\\
$3 \times \{1,1,1,\tau\}$\\
$1 \times \{\tau,\tau,\tau^2\}$\\
$2 \times \{1,1,\tau\}$\\
1 pentagon of edge length 1
} & $\frac{12\tau+7}{12}$ & $-10\bigotimes\bar{\alpha}$ \\
$\bar{T_2}$ & \includegraphics[width=2cm]{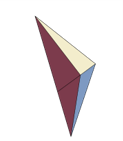} & 4 & 6 & 4 & 
\makecell{
$2 \times \{\tau,1,\tau\}$\\
$2 \times \{\tau,\tau^2,\tau\}$\\
} & $\frac{2\tau+1}{12}$ & $5\tau\bigotimes\bar{\alpha}$ \\
$\bar{T_3}$ & \includegraphics[width=2cm]{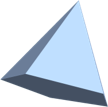} & 6 & 10 & 6 & 
\makecell{
$5 \times \{\tau,1,\tau\}$\\
1 pentagon of edge length 1
} & $\frac{4\tau+3}{12}$ & $5\left(1-\tau\right)\bigotimes\bar{\alpha}$ \\
$\bar{T_4}$ & \includegraphics[width=2cm]{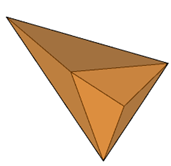} & 5 & 9 & 6 & 
\makecell{
$3 \times \{1,\tau,1\}$\\
$3 \times \{\tau^2,\tau,\tau^2\}$\\
} & $\frac{6\tau+3}{12}$ & $0$ \\ \hline
\end{tabular}
\end{center}
\end{table}
The inflation matrix leads to interesting properties through its right and left eigenvectors \cite{Baake2020}. 
The right eigenvector corresponding to the Perron--Frobenius (PF) eigenvalue $\tau^3$ has components $
(V_{\bar{T}_1}, V_{\bar{T}_2}, V_{\bar{T}_3}, V_{\bar{T}_4})^T,$ and with statistical normalization, it reads $\left(
\frac{1}{2},
\frac{4-\sqrt{5}}{22},
\frac{-5+4\sqrt{5}}{22},
\frac{12-3\sqrt{5}}{22}
\right)^T
\cong
(0.50,\,0.08,\,0.18,\,0.24)^T,$ leading to the relative frequencies of the tiles $\bar{T}_1$, $\bar{T}_2$, $\bar{T}_3$, and $\bar{T}_4$. This implies that the relative frequency of the volume of tile $\bar{T}_1$ is exactly $0.50$. The statistically normalized left eigenvector of $N$ (or equivalently the right eigenvector of $N^T$) is $\left(
\frac{3-\sqrt{5}}{4},
\frac{3-\sqrt{5}}{2},
\frac{3-\sqrt{5}}{2},
\frac{-11+5\sqrt{5}}{4}
\right)^T
\cong
(0.19,\,0.38,\,0.38,\,0.045)^T,$ which gives the relative frequencies of the tiles. In particular, tiles $\bar{T}_2$ and $\bar{T}_3$ each occur with frequency approximately $0.38$, while $\bar{T}_4$ occurs less frequently, at about $0.045$. The PF projection matrix is determined as
\begin{equation}
\begin{alignedat}{1}
\lim\limits_{n \to \infty}{{\tau^{-3n}M}^n=P=\frac{1}{30}}\left(\begin{matrix}\begin{matrix}\begin{matrix}8\tau+1\\2\tau-1\end{matrix}&\begin{matrix}16\tau+2\\4\tau-2\end{matrix}\end{matrix}&\begin{matrix}\begin{matrix}16\tau+2\\4\tau-2\end{matrix}&\begin{matrix}-6\tau+13\\-4\tau+7\end{matrix}\end{matrix}\\\begin{matrix}\begin{matrix}5\\6\tau-3\end{matrix}&\begin{matrix}10\\12\tau-6\end{matrix}\end{matrix}&\begin{matrix}\begin{matrix}10\\12\tau-6\end{matrix}&\begin{matrix}10\tau-15\\-6\tau+18\end{matrix}\end{matrix}\end{matrix}\right),\ \ \ {\ P}^2=P.
\end{alignedat}
\end{equation}
\noindent\hspace*{0.6cm}
The dodecahedron of edge length 1 can be tiled as $d\left(1\right)=3\bar{T_1}+T4$ where the tiles $3\bar{T_1}$ are organized in a threefold symmetric structure and the tile $\bar{T_4}$ is self symmetric under the dihedral group ${W(a}_{2})$. This point will be discussed further in Sec. 5. \newline\noindent\hspace*{0.6cm} 
The new inflation matrix $N$ can also be determined from the following relations. The inflation matrix satisfies the relations when acting on the volume vector (scaled by 12) and the Dehn vector (scaled by $-1/5$),
\begin{subequations}\label{eq:14}
\begin{align}
N \begin{pmatrix}
12\tau + 7 \\
2\tau + 1 \\
4\tau + 3 \\
6\tau + 3
\end{pmatrix}
&=
\begin{pmatrix}
50\tau + 31 \\
8\tau + 5 \\
18\tau + 11 \\
24\tau + 15
\end{pmatrix}, 
N \begin{pmatrix}
2 \\
-\tau \\
\tau - 1 \\
0
\end{pmatrix}
=
\begin{pmatrix}
2\tau \\
-\tau - 1 \\
1 \\
0
\end{pmatrix},\\
\intertext{which can be expressed as a matrix equation provided that $N$ is a non-negative integer matrix:}
N\left(\begin{matrix}\begin{matrix}12&7\\2&1\end{matrix}&\begin{matrix}0&2\\-1&0\end{matrix}\\\begin{matrix}4&3\\6&3\end{matrix}&\begin{matrix}1&-1\\0&0\end{matrix}\end{matrix}\right)
&=
\left(\begin{matrix}\begin{matrix}50&31\\8&5\end{matrix}&\begin{matrix}2&0\\-1&-1\end{matrix}\\\begin{matrix}18&11\\24&15\end{matrix}&\begin{matrix}0&1\\0&0\end{matrix}\end{matrix}\right).
\end{align}
\end{subequations}
The resulting formulation leads to the inflation matrix given in equation (12).


\section{Projection of the MMS tiles from the facets of the Delone cells of \texorpdfstring{$D_6$}{D6}}


One can check from the list of 3D facets of the Delone cells ${W(d}_6)\omega_1$, ${W(d}_6)\omega_5$  and ${ W(d}_{6})\omega_6$ given in Appendix A where the number of tetrahedra in 6D is 1520. Of these, 300 project into trapezoids with edge lengths $\left(1,1,1,\tau\right)$, while the remaining 1220 project into the fundamental tiles $t_i\left(i=1,\ldots,6\right)$, as shown in Table 1. Since the higher dimensional 4D and 5D facets of the Delone cells consist of the 3D tetrahedra,  it is expected that the composite tiles like MS or MMS tiles can naturally be projected from the higher dimensional facets of the Delone cells. In this section we will explore how those composite MS and MMS tiles arise from the projections of the 4D and 5D facets of the Delone cells.\newline\noindent\hspace*{0.6cm}
All the composite tiles can be obtained from the projections of the Delone cells ${W(d}_{6})\omega_5$  and ${W(d}_{6})\omega_6$ as they project into all fundamental tiles however one can get only one composite tile ${T}_3$ from the  projection of the Delone cell ${W(d}_{6 })\omega_1$ only. Indeed, some 5D facets of the Delone cell ${W(d}_{6})\omega_1$ project into the composite tile $T_3$. The Delone cell ${W(d}_{6})\omega_1$ has 12 vertices and it directly projects into an icosahedron in which the composite tile $T_3$ is used in the construction (\cite{Koca2020,Koca2021} for further discussions). The 5D facets of ${W(d}_{6})\omega_1$ are the 5-simplices consisting of 6 vertices and not all of them project onto the composite tile $T_3$. But, for example, the 5-simplex with vertices $l_1,l_2,{-l}_3,{-l}_4,l_5,l_6$ projects into the composite tile $T_3$.
\newline\noindent\hspace*{0.6cm}
Obtaining the rest of the composite tiles from the 4D and 5D facets of the Delone cells ${W(d}_{6})\omega_5$ and ${W(d}_{6})\omega_6$ is more intrigued and needs some care. We will work with the facets of the Delone cell  ${W(d}_{6})\omega_5$ because the facets of the the Delone cell ${W(d}_{6})\omega_6$ are the copies of those of the Delone cell ${W(d}_{6})\omega_5$ as implied by the Dynkin diagram symmetry and can be seen from Appendix A.\newline\noindent\hspace*{0.6cm}
The 4D facets of the Delone cell ${W(d}_{6})\omega_5$ can be treated in two groups. One type of facet is 4-simplex and the other is 4D hemicube. The facet 4-simplices of ${W(d}_{6})\omega_5$ project as those  similar to the 4-simplices of ${W(d}_{6})\omega_1$ so we will not consider such facets. But the 4D hemicube of the Delone cell ${W(d}_{6})\omega_5$ is a new facet and will be considered in some length. We take two typical 4D hemicubes from the orbit ${W(d}_{6})\omega_5$ as shown below,
\begin{subequations}\label{eq:15}
\begin{align}
\frac{1}{2}\left(l_5+l_6\right)+\frac{1}{2}\left(\pm l_1 \pm l_2 \pm l_3 \pm l_4\right)
&\quad \text{with odd $(-)$ signs in the second bracket,} \\
\frac{1}{2}\left(l_5-l_6\right)+\frac{1}{2}\left(\pm l_1 \pm l_2 \pm l_3 \pm l_4\right)
&\quad \text{with even $(+)$ signs in the second bracket.} 
\end{align}
\end{subequations}
The first 4D hemicube (15a) consists of 16 tetrahedral facets and projects onto the composite tile $T_2$. Similarly, the second 4D hemicube (15b) consists also of 16 tetrahedral facets and project to the composite MS tile ${T}_1$. One can check from Appendix A that there are more 4D hemicubes similar to these two classes and all similar 4D hemicubes project in the same manner. Note that (15b) has 8 vertices which is exactly the same number of vertices of the composite tile ${T}_1$. It is then natural to expect that all 4D hemicubes of type (15b) project directly to the tile ${T}_1$. Combining (15a) and (15b) we obtain a 5D hemicube which projects into the composite tiles    
${T}_1,{T}_2$ and ${T}_4$, and when proper combinations are arranged then the MMS tiles follow as $\bar{T_1}={T}_1+{T}_2+{T}_4$ and $\bar{T_4}={T}_2+{T}_4$. There is even a simpler construction of the MMS tile $\bar{T_1}$. We know that the Delone cell ${W(d}_6)\omega_5$  when projected into 3D Euclidean space the vertices split as $32=12+20$ with norm $\left(\frac{\sqrt{3+\sqrt5}}{2}\cong1.14\right)$ and $\left(\sqrt{\frac{3}{5+\sqrt5}}\cong0.64\right)$ respectively showing that an icosahedron embeds a dodecahedron. On the other hand, in the projection of the Delone cell ${W(d}_6)\omega_6$ a dodecahedron embeds an icosahedron. When one considers a subset of 16 vertices of the 32 vertices forming a facet of 5D hemicube then the 16 vertices split as $16=10+6$  in which the 10 vertices form a frustum. When a vertex is removed then the rest of the 9 vertices represent the MMS tile $\bar{T_1}$. This shows us that the composite MMS tiles can be obtained from the projections of the 4D and 5D facets of the Delone cells ${W(d}_{6})\omega_1, { W(d}_{6})\omega_5$ and ${W(d}_{6})\omega_6$.


\section{Threefold symmetric embedding of MMS tiles in dodecahedron}

As we have mentioned in Sec.2 that the orbits ${\ W(h}_{3})v_1,\ {W(h}_{3})v_2$ and ${W(h}_{3})v_3$ represent the vertices of an icosahedron, an icosidodecahedron and a dodecahedron, respectively. Among these, the icosahedron consists of equilateral faces and the icosidodecahedron has the equilateral and pentagonal faces. These two polyhedra with edge lengths $1$ and $\tau$ can be dissected in terms of fundamental tiles ${\ t}_i\left(i=1,2,\ldots,6\right)$ \cite{Koca2021}. But they cannot be dissected in terms of MS or MMS tiles since MS and MMS tiles do not possess equilateral faces. Similarly, the inflated icosahedral tiles or any other inflated tiles having faces with threefold symmetry cannot be dissected in terms of MS or MMS tiles. Therefore, only the dodecahedron can be dissected in terms of MS or MMS tiles. \newline\noindent\hspace*{0.6cm}
In this section we will explain how dodecahedron is dissected in terms of MMS tiles with threefold symmetric embedding. The dodecahedron has threefold symmetry axes through its vertices; this is clear from the invariance of the weight $v_3$ under the group $<R_1,R_2>v_3=v_3$.  As we have already pointed out, in Sec. 3, that the dodecahedron of edge length 1 can be written as $d\left(1\right)=3\bar{T_1}+\bar{T_4}$ including three copies of $\bar{T_1}$ and a centrally symmetric $\bar{T_4}$. As what follows, we will explicitly show that the vertices of dodecahedron split according to the threefold symmetry. For this, one should recall that the generators of the icosahedral group $R_1=r_1r_5,R_2=r_2r_4$ and $R_3=r_3r_6$ transform the orthonormal vectors $l_i$ as 
\begin{equation}
\begin{aligned}
R_1&=r_1r_5:\ l_1\longleftrightarrow l_2;{\ l}_5\longleftrightarrow l_6, \\
R_2&=r_2r_4:\ l_2\longleftrightarrow l_3;{\ l}_4\longleftrightarrow l_5, \\
R_3&=r_3r_6:\ l_3\longleftrightarrow l_4;{\ l}_5\longleftrightarrow-l_6, \\
\end{aligned}
\end{equation}
It is clear that under the threefold symmetry the vectors $l_i$ transform accordingly:
\begin{equation}
\begin{aligned}
R_1R_2:   l_1\longrightarrow l_2\longrightarrow l_3\longrightarrow l_1;  l_4\longrightarrow l_6\longrightarrow l_5\longrightarrow l_4.
\end{aligned}
\end{equation}
The equations in (16-17) imply that the vector ${v}_3$ introduced in equation (7) is invariant under the dihedral group $<R_1,R_2>$  so that the vertices of the dodecahedron besides $\pm{v}_3$ are rotated into each other. \newline
Let us rewrite the equation (7) by substituting  $m_1=m_4=1$ as
\begin{equation}
\begin{aligned}
\frac{1}{2}\left[\left(l_1+l_2+l_3\right)+\left(l_4+l_5+l_6\right)\right]_\parallel=[\omega_6]_\parallel=\frac{1}{\sqrt{2+\tau}}{v}_3.
\end{aligned}
\end{equation}
This is the projection of the vector $\omega_6$ into the space $E_\parallel$. Applying the icosahedral group ${ W(h}_{3})$ on (18), the left-hand side is the set of 20 vectors written in terms of $l_i$ and the right-hand side is the set of vertices of the dodecahedron displayed in (5). Let us classify the vertices as projected pentagons with the sets $\pm A_i$ and $\pm B_i,\left(i=1,2,3,4,5\right)$  given as follows: 
\begin{equation}
\begin{aligned}
A_1&=\frac{1}{2}\left(-l_1+l_2-l_3+l_4-l_5-l_6\right),
B_1=\frac{1}{2}\left(l_1+l_2+l_3+l_4+l_5+l_6\right)\\
A_2&=\frac{1}{2}\left({-l}_1-l_2-l_3+l_4-l_5+l_6\right),
B_2=\frac{1}{2}\left(l_1+l_2-l_3+l_4+l_5-l_6\right)\\
A_3&=\frac{1}{2}\left(-l_1+l_2-l_3-l_4-l_5+l_6\right), B_3=\frac{1}{2}\left(-l_1-l_2-l_3+l_4+l_5-l_6\right)\\
A_4&=\frac{1}{2}\left({-l}_1+l_2+l_3+l_4-l_5+l_6\right),
B_4=\frac{1}{2}\left(-l_1-l_2-l_3-l_4+l_5+l_6\right)\\
A_5&=\frac{1}{2}\left(l_1+l_2-l_3+l_4-l_5+l_6\right), \ \
B_5=\frac{1}{2}\left(-l_1+l_2+l_3-l_4+l_5+l_6\right).      
\end{aligned}
\end{equation}
Face first projection of dodecahedron is shown in Fig. 2.
\begin{figure}[ht] %
\begin{center}
\includegraphics[width=0.4\textwidth]{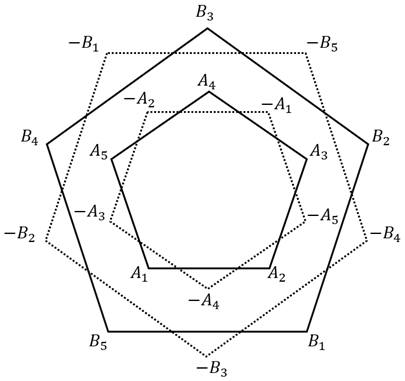} 
\end{center}
\caption{Dodecahedron of edge length 1 with vertices $\pm A_i$ and $\pm B_i$.}
\end{figure}\newline
The nine vertices of one of the tile $\bar{T_1}$ can be chosen as the set
\begin{equation}
\begin{aligned}
({\bar{T_1})}_a= (A_1,{A_2,A_3,A_4,A_5,-B}_1,-B_5,-B4,-B3).
\end{aligned}
\end{equation}
Note that the first five vertices form a pentagonal face of $\bar{T_1}$. The action of the group element  $R_1R_2$ on the vertices in (20) permutes them as follows
\begin{equation}
\begin{aligned}
({\bar{T_1})}_a=\ \left(\begin{matrix}A_1\\A_2\\A_3\\A_4\\A_5\\-B_1\\-B_5\\-B_4\\-B_3\end{matrix}\right),\rightarrow{\bar{T_1})}_b=\left(\begin{matrix}-B_2\\\ \ \ B_4\\-A_5\\-A_1\\\ \ \ B_5\\-B_1\\\ \ A_3\\-B_3\\-A_2\end{matrix}\right),\rightarrow({\bar{T_1})}_c=\left(\begin{matrix}-A_4\\{\ \ B}_3\\-B_5\\\ \ B_2\\-A_3\\-B_1\\-A_5\\-A_2\\-B_4\end{matrix}\right).
\end{aligned}
\end{equation}
Here the vertex $-B_1$ is common for three copies of the tile $\bar{T_1}$; the vertices $-B_3, -B_4, -B_5, -A_2, A_3$ and $-A_5$ are shared two by two among the three sets. Equation (21) should be understood that the tiles  $({\bar{T_1})}_a, ({\bar{T_1})}_b$ and $({\bar{T_1})}_c$ are permuted vertex by vertex so that the symmetry among the  three copies of the tile $\bar{T_1}$ is a dihedral group of order 6. The tile $\bar{T_4}$ has the coordinates, 
\begin{equation}
\begin{aligned}
\bar{T_4}{:\ (\pm B}_1,-B_4,{-B}_3,{-A}_2),
\end{aligned}
\end{equation}
which is invariant under the dihedral group ${W(a}_{2})=<R_1,R_2>$.\newline
Embedding of the tiles $({\bar{T_1})}_a, ({\bar{T_1})}_b$ and $({\bar{T_1})}_c$ in dodecahedron is shown in Fig. 3 by the colors yellow, red and green  respectively and the tile $\bar{T_4}$ by blue.
\begin{figure}[ht] 
\begin{center}
\includegraphics[width=0.27\textwidth]{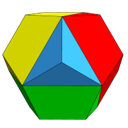} 
\end{center}
\caption{Embedding of the MMS tiles in the dodecahedron.}
\end{figure}\newline\noindent\hspace*{0.6cm}
In the reference \cite{Koca2021} the dodecahedron was tiled with the original MS tiles which included three inner vertices at the intersection of certain MS tiles. With the new definition of the tiles, they are removed so that the dodecahedron tiled with MMS tiles does not possess any inner coordinate.\newline\noindent\hspace*{0.6cm}
There is a similar structure for the projection of the root lattice $A_4$ where the Delone cells project onto the Robinson triangles $\left(1,\tau,1\right),\ (\tau,1, \tau)$ and $\left(1,\tau,1\right)$. They tile the pentagon in a symmetric manner which leaves $(\tau,1, \tau)$ invariant but exchanges two Robinson triangles of type $\left(1,\tau,1\right)$. Here the embedding of the dihedral group ${W(h}_{2})\subset{W(a}_{4})$ plays the same role in 2D tiling as ${W(h}_{3})\subset{W(d}_{6})$ plays in 3D tiling as discussed above.\newline\noindent\hspace*{0.6cm}
We will illustrate the MMS tiling scheme with some examples in which dodecahedral structures are involved. Let us denote the dodecahedra of edge lengths 1, $\tau$ and $\tau^2$ as $d\left(1\right),d\left(\tau\right)$ and $d\left(\tau^2\right)$, respectively. Then they dissect into the MMS tiles as
\begin{subequations}\label{eq:23}
\begin{align}
d\left(1\right)&=3\bar{T_1}+\bar{T_4}, \\
d\left(\tau\right)&=7\bar{T_1}+3\bar{T_4}+8\bar{T_2}+14\bar{T_3},\\
d\left(\tau^2\right)&=7 d\left(1\right)+10\bar{T_1}+50\bar{T_2}+56\bar{T_3}.
\end{align}
\end{subequations}
Note that the dodecahedron $d\left(\tau\right)$ of edge length $\tau$ in (23b) cannot be dissected into the tile of dodecahedron $d\left(1\right)$. Therefore, all tiles $d\left(\tau^n\right),\ n\geq3$  include $d\left(1\right)$ as well as $d\left(\tau\right)$. Figure 4  illustrates the dodecahedron $d\left(\tau^2\right) $ displaying the 7 dodecahedra in a threefold symmetric tiling. Each of $\tau^2 \bar{T_1}$ involves two dodecahedra $d\left(1\right)$ and $\tau^2\bar{T_4}$ contains only one dodecahedron $d\left(1\right)$ where the 7 dodecahedra decompose as $7=2+2+2+1$. This is depicted in Fig. 4. This shows that all inflated dodecahedra $d\left(\tau^n\right),\ n\geq3$ embeds the dodecahedra $d\left(1\right)$ and $d\left(\tau\right)$ in threefold symmetric form. 
\begin{figure}[ht] %
\begin{center}
\includegraphics[width=0.27\textwidth]{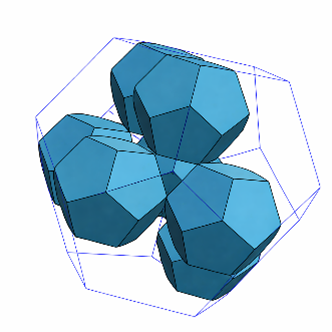} 
\end{center}
\caption{Threefold embedding of 7 dodecahedra $d\left(1\right)$ in the dodecahedron  $d\left(\tau^2\right)$ along with the other tiles (not shown).}
\end{figure}
\section{Discussions}
In this paper we have introduced a modification of the Mosseri-Sadoc tiles abbreviated as the MMS tiles which can be embedded in dodecahedron in a threefold symmetric order. The MMS tiles are obtained by projections of the 4D and 5D facets of the Delone cells of the root lattice D$_{6}$. It is also proved that the dodecahedron inflated by an arbitrary order of the golden ratio can be obtained by taking a subset of the lattice $D_{6}$ without employing the cut-and-project technique. It was also noted that the dodecahedron is the unique icosahedral symmetric polyhedron which embeds the MMS tiles. \newline
The threefold embedding of the MMS tiles is a group-theoretical property that is not only specific to the embedding of icosahedral symmetry $W\left(h_3\right)\subset W\left(d_6\right)$ in the Coxeter–Weyl group $W\left(d_6\right)$. Fivefold symmetric tiling obtained from the projection of the Delone cells of the root lattice $A_{4}$ leads to the twofold symmetric embedding of the Robinson triangles in a pentagon which can be inflated by an arbitrary order by the golden ratio where the dihedral group ${W(h}_{2})\subset{W(a}_{4})$ plays the similar role as ${W(h}_{3})\subset{W(d}_{6})$ plays in the projection from 6D Euclidean space. One can even extend the similar consideration to the 4D tiles tiling the 4D Euclidean space with the non-crystallographic group ${W(h}_{4})$ symmetry embedded in ${W(e}_{8})$, then the 4D tiles can be embedded in the the polytope 120-cell with a ${ W(a}_{3})$ tetrahedral symmetry of order 24. The last one is an academic interest but may naturally arise from these considerations. No one has worked out, so far, the 4D tiling scheme arising from the projection of the lattice $E_8$ with a quasicrystal symmetry ${W(h}_{4})$, an extension of the icosahedral symmetry to the 4D Euclidean space.

\bibliography{iucr} 

@article{AlSiyabi2020,
  author  = {Al-Siyabi, A. and Koca, N. O. and Koca, M.},
  title   = {Symmetry},
  journal = {Symmetry},
  volume  = {12},
  pages   = {1983},
  year    = {2020}
}

@article{Baake2020,
  author  = {Baake, M. and Grimm, U.},
  journal = {Acta Crystallographica Section A},
  volume  = {76},
  pages   = {559--570},
  year    = {2020}
}

@book{Coxeter1973,
  author    = {Coxeter, H. S. M.},
  title     = {Regular Polytopes},
  edition   = {3},
  publisher = {Dover Publications},
  year      = {1973}
}

@incollection{Conway1991,
  author    = {Conway, J. H. and Sloane, N. J. A.},
  title     = {The Cell Structures of Certain Lattices},
  booktitle = {Miscellanea Mathematica},
  editor    = {Hilton, P. and Hirzebruch, F. F. and Remmert, R.},
  publisher = {Springer},
  address   = {New York},
  pages     = {71--108},
  year      = {1991}
}

@book{Conway1999,
  author    = {Conway, J. H. and Sloane, N. J. A.},
  title     = {Sphere Packings, Lattices and Groups},
  edition   = {3},
  publisher = {Springer},
  address   = {New York},
  year      = {1999}
}

@article{Danzer1989,
  author  = {Danzer, L.},
  journal = {Discrete Mathematics},
  volume  = {76},
  pages   = {1--7},
  year    = {1989}
}

@article{Danzer1993,
  author  = {Danzer, L. and Papadopolos, D. and Talis, A.},
  journal = {Journal of Modern Physics B},
  volume  = {7},
  pages   = {1379--1386},
  year    = {1993}
}

@book{DiVincenzo1991,
  author    = {Di Vincenzo, D. and Steinhardt, P. J.},
  title     = {Quasicrystals: The State of the Art},
  publisher = {World Scientific},
  address   = {Singapore},
  year      = {1991}
}

@article{Hann2016,
  author  = {Hann, C. and Socolar, J. E. S. and Steinhardt, P. J.},
  journal = {Physical Review B},
  volume  = {94},
  pages   = {014113},
  year    = {2016}
}

@book{Janot1993,
  author    = {Janot, C.},
  title     = {Quasicrystals: A Primer},
  publisher = {Oxford University Press},
  address   = {Oxford},
  year      = {1993}
}

@incollection{Katz1989,
  author    = {Katz, A.},
  title     = {Some Local Properties of the Three-Dimensional Tilings},
  booktitle = {Introduction to the Mathematics of Quasicrystals},
  editor    = {Jaric, M. V.},
  publisher = {Academic Press},
  address   = {New York},
  pages     = {147--182},
  year      = {1989}
}

@article{Koca2015,
  author  = {Koca, M. and Koca, N. and Koc, R.},
  journal = {Acta Crystallographica Section A},
  volume  = {71},
  pages   = {175--185},
  year    = {2015}
}

@article{Koca2019,
  author  = {Koca, N. O. and Al-Siyabi, A. and Koca, M. and Koc, R.},
  journal = {Symmetry},
  volume  = {11},
  pages   = {2018},
  year    = {2019},
  note    = {arXiv:1811.12176}
}

@misc{Koca2020,
  author = {Koca, M. and Koc, R. and Koca, N. O. and Al-Siyabi, A.},
  note   = {arXiv:2008.00862},
  year   = {2020}
}

@article{Koca2021,
  author  = {Koca, M. and Koc, R. and Koca, N. O. and Al-Siyabi, A.},
  journal = {Acta Crystallographica Section A},
  volume  = {77},
  pages   = {105--116},
  year    = {2021}
}

@article{Kramer1982,
  author  = {Kramer, P.},
  journal = {Acta Crystallographica Section A},
  volume  = {38},
  pages   = {257--264},
  year    = {1982}
}

@article{Kramer1994,
  author  = {Kramer, P. and Papadopolos, Z.},
  journal = {Canadian Journal of Physics},
  volume  = {72},
  number  = {7-8},
  pages   = {408--414},
  year    = {1994}
}

@incollection{Mosseri1982,
  author    = {Mosseri, R. and Sadoc, J. F.},
  title     = {Two and Three Dimensional Non-Periodic Networks Obtained from Self-Similar Tiling},
  booktitle = {The Structure of Non-Crystalline Materials},
  publisher = {Taylor and Francis},
  address   = {London},
  pages     = {137--150},
  year      = {1982}
}

@article{Roth1993,
  author  = {Roth, J.},
  journal = {Journal of Physics A: Mathematical and General},
  volume  = {26},
  pages   = {1455--1461},
  year    = {1993}
}

@article{Papadopolos2000,
  author  = {Papadopolos, Z. and Ogievetsky, O.},
  journal = {Materials Science and Engineering A},
  volume  = {294-296},
  pages   = {385--388},
  year    = {2000}
}

@book{Senechal1995,
  author    = {Senechal, M.},
  title     = {Quasicrystals and Geometry},
  publisher = {Cambridge University Press},
  address   = {Cambridge},
  year      = {1995}
}

@article{Shechtman1984,
  author  = {Shechtman, D. and Blech, I. and Gratias, D. and Cahn, J. W.},
  journal = {Physical Review Letters},
  volume  = {53},
  pages   = {1951--1954},
  year    = {1984}
}

@article{Socolar1986,
  author  = {Socolar, J. E. S. and Steinhardt, P. J.},
  journal = {Physical Review B},
  volume  = {34},
  pages   = {617--647},
  year    = {1986}
}

@article{Steurer2004,
  author  = {Steurer, W.},
  journal = {Zeitschrift für Kristallographie},
  volume  = {219},
  pages   = {391--446},
  year    = {2004}
}

@article{Tsai1994,
  author  = {Tsai, A. P. and Niikura, A. and Inoue, A. and Masumoto, T. and Nishida, Y. and Tsuda, K. and Tanaka, M.},
  journal = {Philosophical Magazine Letters},
  volume  = {70},
  number  = {3},
  pages   = {165--175},
  year    = {1994}
}

@article{Tsai2008,
  author  = {Tsai, A. P.},
  journal = {Science and Technology of Advanced Materials},
  volume  = {9},
  pages   = {013008},
  year    = {2008}
}

\newpage

\appendix
\section*{Appendix A: Facets of the fundamental polytopes of $D_6$}

\addcontentsline{toc}{section}{Appendix A: Facets of the fundamental polytopes of \texorpdfstring{$D_6$}{D6}}

\begin{table}[ht]
\centering
\caption{Facets of the fundamental polytopes of $D_6$, where $N_i$ $(i=0,1,\ldots,5)$ denotes the number of facets*.}
\renewcommand{\arraystretch}{1.7}
\small
\begin{tabular}{|c|c|c|c|c|c|c|}
\hline
Orbits & $N_0$ & $N_1$ & $N_2$ & $N_3$ & $N_4$ & $N_5$ \\
\hline
${W(D_6)\omega_1}$ & 12 & 60 & 160 & 240 (tetrahedron) & 192 (4-simplex) & 64 (5-simplex)\\  \hline
${W(D_6)\omega_2}$ & 60 & 480 & 1120 & 
\makecell{$1200 = $\\ $240$ (octahedron)\\$+\ 960$ (tetrahedron)} & 576 (4-simplex) &
\makecell{$76 =$ \\ $64$ (ambo-5-simplex)\\$+\ 12$ (5-hemicube)} \\  \hline
${W(D_6)\omega_3}$ & 160 & 1440 & 2880 & 
\makecell{$2160 =$ \\ $960$ (octahedron)\\$+\ 1200$ (tetrahedron)} & \makecell{$636 =$ \\ $576$ (4-simplex)\\$+\ 60$ (4-hemicube)} &
\makecell{$76 =$\\ $64$ (ambo-5-simplex) +\\$12$ (ambo-5-hemicube)} \\  \hline
${W(D_6)\omega_4}$ & 240 & 1920 & 3200 & 
\makecell{$2080 = $ \\ $960$ (octahedron)\\$+\ 1120$ (tetrahedron)} & \makecell{$636 = $ \\   $576$ (4-simplex)\\$+\ 60$ (24-cell)} &
\makecell{$76 =$ \\  $64$ ($3^{rd}$\ ambo-5-simplex) +\\$12$ ($2^{nd}$\ ambo-5-hemicube)} \\  \hline
${W(D_6)\omega_5}$ & 32 & 240 & 640 & 
640 (tetrahedron) & \makecell{$252 =$ \\ $192$ (4-simplex)\\$+\ 60$ (4-hemicube)} &
\makecell{$44 =$ \\ $32$ (5-simplex) +\\$12$ (5-hemicube)} \\  \hline
${W(D_6)\omega_6}$ & 32 & 240 & 640 & 
640 (tetrahedron) & \makecell{$252 =$ \\ $192$ (4-simplex)\\$+\ 60$ (4-hemicube)} &
\makecell{$44 =$ \\ $32$ (5-simplex)+\\$12$ (5-hemicube)} \\ 
\hline
\end{tabular}
\end{table}
\vspace*{0.5em}
{\leftskip=2em
\noindent\textbf * Names of the polytopes follow \cite{Conway1991}.}
\end{document}